# Unraveling atomic-resolution valence electron energy-loss spectroscopic imaging in a single-crystal $CaNb_2O_6$

Sz-Chian Liou,[1,*,†] Xiang-Lin Huang,[2,*] Vladimir P. Oleshko,[3,*] I-Ching Lin,[4] Yin-Ping Lan,[5] Hsin-An Chen,[5,6] Guo-Jiun Shu,[2,6,‡]

**AFFILIATIONS**

1. Institute of Functional Materials and Devices, Lehigh University, Bethlehem, PA 18015, USA.
2. Institute of Mineral Resources Engineering, National Taipei University of Technology, Taipei 10608, Taiwan.
3. Scientific Consulting, Research & Technical Service, Gaithersburg, MD 20879, U.S.A.
4. Condensed Matter Physics and Materials Science Department, Brookhaven National Laboratory, Upton, New York 11973, U.S.A.
5. Department of Materials and Mineral Resources Engineering, National Taipei University of Science and Technology, Taipei 10608, Taiwan.
6. Institute of Materials Science and Engineering, National Taipei University of Technology, Taipei 10608, Taiwan.

*These authors contributed equally to this work.

[†]Contact author: szl223@lehigh.edu
[‡]Contact author: gjshu@mail.ntut.edu.tw

# ABSTRACT

Despite advancements in electron optics and spectrometer design over the past twenty years, atomic-resolution valence-electron energy-loss spectroscopy imaging remains challenging due to the delocalization of inelastic electron scattering. In this study, we used an energy-filtered spectrometer equipped with a hybrid-pixel direct electron detector and spherical aberration-corrected scanning transmission electron microscopy to analyze many-electron excitations and interband transitions in a single-crystal calcium niobate, $CaNb_2O_6$, with spatial resolution ranging from the nanometers to the atomic scale. In the low-loss region above the bandgap at about 3.8 eV, we observed volume plasmons, around 6 eV and 15 eV energy loss, as well as a mix of strongly correlated plasmons and excitons, known as plexcitons, at approximately 7.3 eV energy loss. Additionally, we employed an on-axis EELS setup for atomic-resolution zero-loss peak (ZLP) imaging and visualized energy- and atom-resolved images of plexcitons and VPs, which showed contrast reversal relative to high-angle annular dark-field images. To investigate elastic contrast preservation, we also analyzed the effect of the collection angle and minimized its influence to produce delocalized VP images. In fact, the ZLP and VEELS images obtained using the weak-beam setup demonstrate that, in both cases, the contrast resembles Z-contrast. Moreover, we found that [$NbO_6$] octahedra directly contributed to the lateral maps of interband transitions in the range from 3.2 eV to 3.5 eV energy loss. These findings demonstrate that Cs-STEM-EELS, which examines atomic-scale contrast associated with low-energy losses, can be a powerful tool for visualizing the structure, bonding, and electronic properties of complex crystalline nanostructures, including individual atomic sites, interstitial sites, and point defects.

## I. INTRODUCTION

Over the past few decades, electron energy-loss spectroscopy (EELS) has become an essential tool for analyzing detailed electronic-structure information, including phonons excited by lattice vibrations, various low-loss excitations, transitions between the valence and conduction bands, and core-loss excitations from inner-shell electrons [1-3]. Significant improvements in energy resolution enabled by electron monochromators now allow one to investigate the phonon region with EELS [3]. With commercialized direct electron detectors [4,5] that improve signal-to-noise (S/N) ratios, along with adjustments of post-specimen-lens settings [6-8] and innovative spectrometer designs [9], EELS's detection efficiency has greatly increased, extending the accessible energy-loss range to higher energies—more than a few thousand eV - approaching 10 keV to 30 keV energy losses. This range now rivals X-ray absorption spectroscopy (XAS) using synchrotron radiation [6-9]. Meanwhile, with major advancements in spherical aberration correction, Cs-STEM-EELS equipped with an electron probe of about 70 pm or smaller has achieved atomic resolution, opening new opportunities for atomic-scale mapping.

However, in addition to the picometer-scale electron probe, the inelastic scattering angle ($\theta_E$, roughly $\sim\frac{\Delta E}{2E_0}$, where $E_0$ is the energy of incident electrons) and its delocalization must be considered for the realization of atomic-resolution EELS mapping. At an accelerating voltage of 200 keV, $\theta_E$ varies from 0.0025 mrad to 0.125 mrad in the low-loss region from 1 eV to 50 eV, and up to 2.5 mrad in the core-loss region from 55 eV to 1000 eV. Consequently, according to the formula $L_{50}\approx \frac{0.5\lambda}{(\theta_E)^{3/4}}$, where λ is the electron wavelength, the delocalization length ranges from 20 nm at 1 eV to nearly 1 nm at 50 eV, then decreases to 0.1 nm at 1000 eV [1]. In the core-loss region, EELS mapping can therefore approach atomic resolution as energy loss increases, and atomic-resolution core-level elemental mapping has been realized and used to analyze bonding and coordination environments at atomic columns [10]. Due to constraints induced by delocalization in the phonon and low-loss regions, the atomic-resolution limit, however, is evidently exceeded. Instead of using central Bragg reflections, a weak-beam geometry, or the so-called off-axis EELS setup, was proposed [11], in which inelastic scattering can be localized by displacing the collecting EELS entrance aperture so that it no longer overlaps any Bragg reflections. The inelastic scattering should then be confined to the probe wave function [11]. Thus, atomic-resolution phonon mapping employing monochromated Cs-STEM-EELS in a weak-beam geometry, also known as the off-axis EELS setup, which uses large-angle Kikuchi bands as a phonon spectral source, has been demonstrated recently [12].

On the contrary, atomic-resolution zero-loss imaging and valence EELS (VEELS) imaging have been demonstrated using energy-filtered TEM (EFTEM) [12-16], while atomic-resolution VEELS imaging with STEM has shown inconsistent results for the same monolayer graphene [16-18]. These STEM-EELS studies revealed only a delocalized signal in pristine and doped monolayer graphene [17, 18]. This likely results from the tail of the strong zero-loss peak (ZLP) and the limited detection efficiency of spectrometer cameras, both of which hinder localized atomic-resolution VEELS imaging. Conversely, atomic-resolution plasmon imaging of monolayer graphene has been demonstrated using EFTEM, with contrast that was varying with defocus [16]. In contrast, atomic-resolution VEELS imaging has been successfully performed on thin crystalline silicon samples and on a $(BaTiO_3)_8/(SrTiO_3)_4$ heterostructure using a similar spectrometer [19,20]. These findings indicate that the type of crystalline material can be as important as the instrumentation. Interestingly, both zero-loss and volume plasmon (VP) images show contrast reversal relative to HAADF signals [5,19,20], suggesting that the maximum intensity does not always correspond to atomic columns, unlike phonon [12] and core-loss maps [10]. These contrast variations may be due to the preservation of elastic contrast, where inelastic intensity is more affected by interference of scattered electrons during elastic propagation rather than by the distribution of inelastically scattered electrons [14,16,21-23]. Several methods have been proposed to deal with this problem, i.e., energy-filtered imaging STEM (EFISTEM) [22]; increasing the collection angle beyond the Bragg reflection angle to make the inelastically scattered electrons incoherent [11,23]; and deconvolving incoherent bright-field (IBF) images to obtain an elastic correction map [20]. However, the effects of collection angles and the weak-beam setup on atomic-resolution VEELS imaging have not been discussed in detail. Therefore, exploring atomic-resolution VEELS imaging can be valuable not only for mono-elemental, light-atom materials with covalent bonds but also for heteroatomic crystals such as mixed-element oxides, which feature atomic columns of higher-atomic-number elements and partially polarized ionic bonds.

Orthorhombic $CaNb_2O_6$ columbite (abbreviated as CNO), used in this study, is widely recognized as a promising material for various key nanotechnology applications [24-27]. These include potential uses in holography and as a laser host material; it shows strong blue luminescence centered at 465 nm when excited with ultraviolet (UV) light [24-26]; it functions as a microwave dielectric material due to its spontaneous polarization in the orthorhombic crystal symmetry (*Pbcn*) [26]; and it operates in UV photodetectors within optoelectronics because of its high absorption in the UV region below 300 nm and transparency from 400 nm to 1000 nm in the visible to infrared range [27,28]. CNO is known for its luminescence, which results from electronic transitions between the O and Nb bands within the $[NbO_6]$ octahedral units in its anisotropic crystal structure. Recently, ultrathin, non-layered single-crystal CNO nanosheets synthesized via chemical vapor deposition using edge-seeded heteroepitaxy and slow kinetic growth have been used in nanoscale two-dimensional (2D) field-effect transistors (FETs) [27]. These as-grown 2D CNO nanosheets demonstrated excellent crystalline quality, a thickness-independent high dielectric constant ($\kappa$, approximately 16), and a large breakdown field strength of about 12 MV/cm. The $MoS_2$/CNO heterostructure FETs fabricated with these nanosheets exhibited an on/off ratio exceeding $10^7$, negligible hysteresis, and a subthreshold swing of 61 mV/dec, paving the way for scalable, non-layered 2D dielectrics in electronics. Additionally, the asymmetric crystal structure of CNO offers opportunities to design innovative nanoelectromechanical system (MEMS) resonator devices with directional mechanical and thermal properties for potential uses in computation, sensing, and quantum information [28].

Consequently, by selecting an appropriate low-index zone axis, we can exploit the structural anisotropies of the CNO to probe the entire spectral range of low-energy excitations with atomic-resolution VEELS. From the crystallographic projection along the [001] zone axis shown in Fig. 1(a1), it is clear that niobium (Nb) cations are coordinated by six oxygen (O) atoms, forming [$NbO_6$] octahedral units. These units connect through shared edges along the *b*–*c* planes to form nanochains of [$NbO_6$] octahedra. These chains share corners and link to negatively charged double layers, the main structural units, which are further separated by layers of $Ca^{2+}$ ions acting as interlayer counterions [26]. Therefore, along the [001] zone axis, the columns of [$NbO_6$] octahedra and calcium (Ca) atoms are more distinguishable than along the [100] zone axis, as shown in Fig. 1(a2). By positioning an electron probe at different atomic columns and recording the atomic-column-dependent VEEL spectra, we can analyze specific features in the low-loss spectrum and visualize their distribution across the surface in real space. Although CNO exhibits excellent optical properties, systematic studies of its electronic excitations in the low-loss region accessible by optical spectroscopy remain limited. Additionally, the low-loss energy range up to 50 eV, which exceeds the detection capabilities of optical spectroscopy in the visible spectrum, has not yet been explored using EELS.

In this study, using CNO single crystals grown along the [001] zone axis, we report on atomic-resolution VEELS images acquired at different collection angles and under weak-beam conditions. We employed a Cs-STEM and a CEOS energy-filtering and imaging device (CEFID) [29], equipped with a highly sensitive hybrid-pixel Dectris ELA direct electron detector [4], to investigate the crystalline and electronic structures of CNO. Compared with previous energy-filtering imaging systems, the CEFID design enabled us to further reduce higher-order aberrations and minimize geometric and chromatic distortions, thereby achieving lower non-isochromaticity with improved energy resolution [7,8,29]. The electronic structures were examined at the nanometer scale using VEELS in an *aloof* setup, with atomic-scale spatial resolution to collect atomic-column-dependent VEELS signals. We demonstrate that the atomic-resolution zero-loss images and VEELS images obtained under on-axis conditions exhibit contrast reversal relative to the HAADF images due to the preservation of elastic contrast, while those images display the same contrast as the HAADF images using the off-axis conditions.

## II. METHODS

CNO powders were produced through a solid-state reaction by mixing raw $CaCO_3$ (99.9%, Aldrich) and $Nb_2O_5$ (99.9%, Aldrich) powders in stoichiometric ratios. The mixture was calcined in air at 1250 °C for 24 hours to obtain single-phase CNO powders, confirmed by XRD. These powders were ground and pressed into cylindrical rods, 6 mm in diameter and 100 mm long, using about 100 MPa of isostatic pressure to create feed rods. The rods were sintered at 1300 °C for 24 hours in air, sliced into approximately 20 mm segments to serve as initial seed crystals, and then transferred to an optical floating-zone furnace to grow a CNO single crystal.

A Bruker D8 Phaser X-ray diffractometer with a Cu $K\alpha$ source was used to perform structural analysis and phase identification of the CNO powders prepared by crushing a small piece of a single-crystal CNO. The data were collected at room temperature with a step size of 0.02 degrees and a scanning speed of 1 second per step. The collected XRD patterns were further analyzed using the Rietveld method with the DIFFRAC.TOPAS program package.

High-quality TEM specimens were prepared using a focused ion beam (FIB, Thermo Fisher Scios), following the procedure detailed elsewhere [8]. Final cleaning involved argon ion milling (Fischione 1040 NanoMill) to remove ion beam damage and surface amorphous layers. To prevent carbon contamination, the TEM specimens were then plasma cleaned before STEM-EELS experiments. The microstructures and electronic structures of CNO specimens were analyzed with a JEOL-ARM 200 CF operated at 200 kV. This instrument features a cold field emission gun (CFEG), CEOS ASCOR probe aberration corrector, JEOL Dry EDS detector (model SD 100 GV, with a collection solid angle of 0.8 steradian), and a CEFID system with a hybrid-pixel Dectris ELA detector. During Cs-STEM experiments, the following parameters were used: a 70 pm diameter electron probe, a 25 mrad convergent semi-angle ($\alpha$), and a collection semi-angle for HAADF ranging from 78 mrad to 180 mrad. EELS collection semi-angles ($\beta$) of 20, 25, and 50 mrad were selected to examine the impact of collection angle on VEEL signal intensity. Additionally, the weak-beam (off-axis) EELS setup [12] was employed to compare signal-intensity variations between on-axis and weak-beam (off-axis) collection geometries. For EELS measurements, the energy resolution remained at 0.7 eV throughout the STEM-EELS experiments, with non-isochromaticity kept below 70 meV. Parameters for STEM-EEL spectrum imaging (SI) datasets included a scan step of 0.051 nm per pixel and an acquisition dwell time of 5 ms per pixel. Since the total acquisition time for each data cube was under 1 minute, drift correction was unnecessary. Both Cs-STEM HAADF images and EEL spectra were acquired using a JEOL FEMTUS platform [7,8]. The recorded EELS data were converted to Gatan DigitalMicrograph (DM) format for further processing with the Gatan Microscopy Suite (Gatan-AMETEK) [1,8], including deconvolution of raw spectra to derive single-scattering distributions and Kramers-Krönig analysis (KKA) [1,8].

Calculations of the electronic band structure of CNO were carried out from first principles using the Vienna *ab* initio Simulation Package (VASP) based on density functional theory (DFT) [30]. The generalized gradient approximation (GGA) with the revised Perdew-Burke-Ernzerhof exchange-correlation functional for solids (PBESOL) [31] was used for the exchange-correlation functional, along with projector augmented-wave (PAW) pseudopotentials [32]. The electronic structure was fully optimized until all forces fell below $10^{-3}$ eV/nm. A convergence criterion of $10^{-8}$ eV was set for the self-consistent field cycle of the electronic wave functions. A plane-wave basis set cutoff energy of 500 eV was employed. After minimization, the partial density of states (PDOS) of CNO was calculated. Γ-centered k-point meshes were used in all calculations, with a k-point spacing of 1.5 $nm^{-1}$ for PDOS.

## III. RESULTS AND DISCUSSION

Figure 1(b) shows an XRD pattern of CNO powders, created by crushing and grinding a single-crystal CNO ingot [see the inset in Fig. 1(b)]. For comparison, the XRD data for columbite, a *Pbcn* space group [(International Crystal Structure Database (ICSD) card No. 15208) [26]], and Rynersonite, which is isostructural with $CaTa_2O_6$ and has a *Pnma* space group (ICSD card No. 148209) [33], are displayed in Fig. 1(b) by blue and red curves, respectively. The experimental XRD pattern clearly indicates that the CNO sample used in this study is a single orthorhombic phase free of macroscopic impurities, secondary phases, or precipitates. The refined CNO structural details are summarized in Table 1. The refinement $R_{wp}$ value was 8.36%, falling within the 5-8.5% range, which is typically used to indicate high-quality powder X-ray diffraction data [34] for powders prepared by crushing single crystals. The experimental XRD pattern matches the simulated pattern of columbite (ICSD card No. 15208), which belongs to the space group *Pbcn*. The refined lattice constants of the unit cell are: $a$ = 1.4981(2) nm, $b$ = 0.5749(7) nm, and $c$ = 0.5224(6) nm. The Ca atoms occupy the Wyckoff 4c site, while the Nb atom and three different O atoms occupy the 8d sites.

Figure 1(c) displays a representative energy-dispersive X-ray spectrum from the CNO specimen, showing several characteristic X-ray lines, including the prominent Ca $K$-series at 3.69 keV ($K\alpha1$) and 4.01 keV ($K\beta1$), the Nb $L$-series at 2.17 keV ($L\alpha1$) and 2.26 keV ($L\beta1$), the O $K\alpha$ line at 0.53 keV, and the minor Ca $L\alpha$ at 0.34 keV. The X-ray maps for Ca, Nb, and O, shown in yellow, cyan, and red, respectively [see inset in Fig. 1(c)], confirm a uniform distribution of the elements within the specimen.

Figures 1(d1) and 1(e1) show the experimental selected-area electron diffraction (SAED) patterns, where Bragg reflection conditions for the space group *Pbcn* of CNO are: 0*kl*, where *k* is even; *h*0*l*, where *l* is even; *hk*0, where *h* + *k* is even; *h*00, where *h* is even; 0*k*0, where *k* is even; and 00*l*, where *l* is even. The absence of additional Bragg spots supports the XRD results that the CNO sample used in this study exhibited high crystallinity and phase purity. The zone axes of the SAED patterns in Fig. 1(d1) and 1(e1) can be indexed in agreement with XRD results as [001] and [100], respectively, based on the calculated *d*-spacings of the diffraction spots marked with red circles. Corresponding HAADF images of the CNO sample oriented along the [001] and [100] zone axes are shown in Fig. 1(d2) and 1(e2), respectively. The inset in Fig. 1(d2) presents a crystallographic projection showing the positions of Nb and Ca atoms. The HAADF image in Fig. 1(d2) demonstrates a clear contrast between the Nb and Ca atomic columns since the HAADF signal intensity is proportional to the atomic number $Z$, a phenomenon known as $Z$-contrast [8,23]. It displays well-separated Nb atomic columns ($Z_{Nb}$= 41) with brighter contrast and Ca atomic columns ($Z_{Ca}$= 20), which are situated between two Nb layers, perfectly matching the crystallographic projection along the [001] zone axis shown in Fig. 1(a1) and the inset in Fig. 1(d2), overlaid with the HAADF image. To verify the assignment of the Nb and Ca atomic columns, we simulated HAADF images of the CNO sample using a multislice technique with xHREM$^{TM}$ software (HREM Research Inc., Japan). The sample thickness in Fig. 1(d2) was 35 nm, with a ± 2 nm uncertainty (or 70 cells; cell, $c$ = 0.52244 nm), measured from the VEELS spectrum in Fig. 2(a) using a Fourier-log ratio technique [1,8]. To include phonon electron scattering (or thermal diffuse scattering, TDS) in the HAADF simulation, we used the Debye-Waller (DW) factors listed in Table 1 for CNO. The simulated HAADF image [inset in Fig. 1(d2)] shows the same contrast, i.e., brighter Nb atomic columns and lighter Ca atomic columns, with no contrast for O atomic columns. Conversely, the HAADF image along the [100] zone axis in Fig. 1(e2) shows a uniform zigzag arrangement. The projected spacing between the Nb and Ca atomic columns along the [100] zone axis is less than 40 pm, which is smaller than the used electron probe size of 70 pm. This makes it difficult to resolve the contrast between Ca and Nb atomic columns, as seen in the simulated HAADF inset in Fig. 1(e2). Since the [$NbO_6$] octahedral and Ca atomic columns can be distinguished along the [001] zone axis, we used a 70 pm electron probe to position it at different atomic columns and record the atomic-column-dependent VEEL spectra.

To investigate volume- and surface-induced resonances, we used a precise *aloof* setup that involved sequentially shifting the electron probe from the bulk interior to the edge of the specimen at grazing incidence (i.e., $b$ = 0 nm), then to a few nanometers away from the specimen edge. The electron probe was initially located inside the material, as indicated by the purple circle in the HAADF image in Fig. 2(c), and the corresponding VEEL spectrum is shown as the purple curve in Fig. 2(a). Above the optical bandgap ($E_g$) discussed below, the energy-loss spectrum displayed several peaks at approximately 7.3 eV and 14.9 eV, as well as a broad feature at 24.7 eV. Spectral features at higher energy losses of about 34.8 and 38.8 eV were attributed to minor Ca $M_{2,3}$-edges caused by transitions of Ca 3*p* electrons to unoccupied 3*d* states, while the spectral feature at around 45.9 eV energy loss was assigned to the minor Nb $N_{2,3}$-edge related to transitions of Nb 4*p* electrons to unoccupied 4*d* and 5$s$ states. Additionally, a shoulder, marked by a black arrow in Fig. 2(a), at an energy loss of approximately 6 eV was observed.

To interpret the VEEL spectra of CNO crystals, we used the framework of the macroscopic complex dielectric function $\varepsilon(\omega) = \varepsilon_1(\omega) + i\varepsilon_2(\omega)$ [1,2,8,15]. The dielectric function $\varepsilon(\omega)$ of the single-crystal CNO, obtained from the VEEL spectrum [a purple curve in Fig. 2(a)] through a KKA [1,2,8,15], is shown in Fig. 2(b) with its real part, $\varepsilon_1(\omega)$, in gray, and its imaginary part, $\varepsilon_2(\omega)$, in red. Since the condition $\varepsilon_1 = 0$ with a positive slope and a small $\varepsilon_2$ value can fulfill volume plasmon (VP) resonance, the peaks near 6 eV and 14.9 eV in Fig. 2(a) are interpreted as VPs, which correspond to maxima in the volume energy loss function (ELF) $\propto \mathrm{Im}\left\{\frac{-1}{\varepsilon(\omega)}\right\}$. Additionally, the shoulder around 7.3 eV and the broad feature near 24.7 eV are attributed to interband transitions, as indicated by the absorption $\varepsilon_2$ curves near these energies in Fig. 2(b).

The inset in Fig. 2(b) shows a magnified section of the purple spectrum in Fig. 2(a), in the energy-loss range of 2.9 to 5.5 eV, illustrating the measurement of the optical bandgap $E_g$. The $E_g$, approximately 3.8 eV ± 0.05 eV (1.32% relative uncertainty), was calculated using a previously described linear fitting method [15]. The parabolic spectral feature observed above $E_g$ [see Fig. 2(a) and the inset in Fig. 2(b)] indicates an indirect transition in CNO, consistent with reported DFT calculations [35,36]. The valence band maximum is located between the Γ and Z points, while the conduction band minimum is at the Γ point [36]. A blue bracket within the $E_g$ region highlights an additional broadened feature between 3.2 eV and 3.5 eV in the energy-loss spectrum, previously observed in the ultraviolet-visible (UV-Vis) spectra of nanocrystalline CNO [37], and likely represents an intrinsic electronic excitation.

Considering the DFT-calculated partial density of states (PDOS) in Fig. 2(d), the valence band maximum (VBM) is localized from -5 eV to the $E_F$ (energy = 0 eV) and is mainly dominated by hybridization between the O-2$p$ states (olive curve) and the Nb-4$d$ states (pink curve), with a small contribution from the Nb-3$p$ states (cyan curve). The contribution of the Ca states in the VBM region was neglected because it was quite small. The hybridization of the O-2$p$ and Nb-4$d$ states indicates the covalent character of the Nb-O bond in the $NbO_6$ octahedra. The peaks at the bottom of the valence band, between -20 eV and -15 eV, are primarily contributed by the O-2$s$ states (yellow curve) and the Ca-3$p$ states (red curve). In contrast, the conduction band minimum (CBM) between 4 eV and 6 eV is mainly dominated by O-2$p$ and Nb-4$d$ states. The deep conduction band from 7 eV is primarily influenced by the mixing of Ca-3$d$, Nb-4$d$, O-2$p$, and O-3$s$ states (green curve). From the PDOS, it can be deduced that hybridization between the O-2$p$ and Nb-4$d$ states contributes to the VEEL spectral features above the bandgap. Conversely, the broadened spectral features in the 20-30 eV energy-loss range were assigned to hybridized O-$s$ and Ca-$p$ states. Furthermore, the low-intensity feature at an energy loss of about 3.2 to 3.5 eV, dominated by O-2$p$ states and Nb-3$p$ and/or 4$d$ states, was attributed to the [$NbO_6$] octahedra.

To investigate surface-induced resonances, the electron probe was positioned at distances of $b$ = 4 nm and $b$ = 10 nm from the specimen edge [see the color circles in the HAADF image in Fig. 2(c)]. The acquired VEEL spectra are shown in Fig. 2(a) by the red and orange curves, respectively. As observed, the oscillator strengths of the spectral peaks at energy losses of approximately 6 eV, 7.3 eV, and 14.9 eV decreased as $b$ increased. Additionally, the peaks shifted to energy losses of about 5 eV, 6.3 eV, and 10.5 eV, respectively, indicating their surface-excitation character and suggesting they are surface plasmons (SPs) due to the negative $\varepsilon_1$ values [1,2,8,15]. Indeed, the peak at an energy loss of 7.3 eV, previously assigned to either an interband transition or an exciton, also exhibits surface-excitation behavior. Therefore, based on the criterion of $\varepsilon_2 > \varepsilon_1 > 0$ [8,15], the spectral peak at around 7.3 eV can be interpreted as a surface exciton polariton (SEP), representing a collective oscillation of delocalized excitons at the CNO surface. Furthermore, this spectral feature in the 5 to 7 eV energy-loss range may arise from the mixing of plasmon resonances and excitons, forming so-called plexciton excitations [38,39].

Using selected energy windows within an integrated *aloof* and STEM-EELS-SI setup, one can image real-space distributions of specific features in the VEEL spectra of CNO. Figure 2(c) shows the HAADF image and VEELS intensity maps of the spectral features, acquired simultaneously with a 2 eV window at ZLP, 7 eV, and 15 eV energy losses, respectively. The maximum intensity of these spectral features corresponds to the spatial distribution of the plexciton states for energy losses from 5 to 7 eV, revealing a decay of the evanescent wave field into the vacuum and exhibiting SP-like excitations. In contrast, the maximum intensity of the VP at about 15 eV indicates strong localization within the bulk, clearly confirming its volume-resonance nature. This qualitatively agrees with the classical approximation for a delocalization width containing 50% of the scattering electrons, $L_{50} \approx \frac{0.5\lambda}{(\theta_E)^{3/4}}$ [1,11]. At lower energy losses, from 5 eV to 7 eV, smaller $\theta_E$ values can increase $L_{50}$ to 4.6 nm at an accelerating voltage of 200 kV [1,11]. According to this model, due to the long-range electrostatic interaction between the incident electron probe and the sample, the VEELS intensity map should be limited to nanometer-scale resolution, and atomic-resolution VEELS is generally not achievable. However, several reports [5,19,20] along with the results of this work, suggest that atomic-resolution VEELS of a crystalline specimen can still be realized using a 70 pm electron probe in aberration-corrected Cs-STEMs to confine the electron probe along atomic columns.

Positioning such electron probe at the Nb and Ca atomic columns, the O atomic columns between two Nb layers, and the interstitial sites between two Ca columns, marked by the orange, blue, red, and gray circles, respectively, in a HAADF image in Fig. 3(b), allowed us to acquire corresponding ZLP and VEEL spectra using a standard on-axis EELS setup [see schematic inset in Fig. 3(a1) and spectra in Figs. 3(a1)-(a4) shown with the same color coding]. Conducting such analyses, we examined several SI datasets, each containing at least 64, 32, 128, and 32 atom positions for Nb, Ca, and O atomic columns, and interstitial sites, respectively. We calculated both average spectra and ranges of variation for each spectrum shown in Figs. 3(a1)-(a4). We observed a significant decrease in position-dependent ZLP and VEELS intensities in the following order: the interstitial site > the O column between two Nb layers > the Ca column > the Nb column, as shown in Figs. 3(a1)-(a4). The intensities decreased by about 20% along this sequence. It was, however, difficult to assign the O atomic columns due to weak contrast in both HAADF and ZLP images, resulting in a wide range of deviations and greater uncertainty in their positions relative to the Nb and Ca atomic columns and interstitial sites. The results indicate that intensity diminishes with increasing atomic number $Z$ due to stronger interactions between the incident probe and the CNO crystal. In fact, the ZLP image in Fig. 3(c1) shows a spatial resolution similar to that of the HAADF image in Fig. 3(b), but exhibits contrast reversal relative to the HAADF images. Considering the geometry of the on-axis EELS setup, the electron source can be described by the central Ronchigram, with a convergent semi-angle ($\alpha$ = 25 mrad) that overlaps many Bragg reflection discs. The electron beam passes through the inner entrance hole of the ADF detector and continues toward the entrance aperture of the EEL spectroscope. With the bright-field (BF) detector inserted, this setup can also be used to acquire BF STEM images. A similar decrease in the signal intensity with increasing $Z$ was observed in plexciton and VP images shown in Fig. 3(c2) and 3(c3), respectively. The reversed contrast, similar to that of the high-angle BF (HABF) image [19], causes the Nb and Ca atomic columns to appear dark, while the O atomic column appears bright [see the red circles in the schematic atomic arrangements inset in Figs. 3(c1)-(c3)]. In this way, VEELS not only demonstrated atomic resolution but also revealed a maximum-intensity resonance localized around O and interstitial sites, unlike the diffuse maps reported for monolayer graphene [17,18].

Since central Ronchigrams with an $\alpha$ = 25 mrad contain many overlapping Bragg reflections, this results in interference among them, leading to the appearance of phase contrast in BF images [23]. Furthermore, in the on-axis EELS setup, the ZLP image is formed similarly to the energy-filtered BF image, thereby preserving the elastic contrast. To reduce the influence of elastic contrast, the collection angle can be increased beyond the Bragg reflection angle [23]. Using this approach, we kept the constant at $\alpha$ = 25 mrad and recorded BF images at different collection angles, while also acquiring STEM-EELS SI datasets as shown in Figs. 4. We used a set of collection angles that were slightly less than, equal to, and twice the $\alpha$ = 25 mrad. This was done because small collection angles can often produce interference contrast features, causing noticeable differences between experimental and simulated BF images. Inspection of the recorded BF image integrated up to collection angles of 20 mrad in Fig. 4(a1) revealed atomic resolution with dark contrast at the Nb atomic columns (cyan circles), gray contrast at the Ca atomic columns (orange circles), and bright contrast at the O atomic columns (red circles), as well as interstitial sites. The distance between neighboring Nb atomic columns appeared very small, with dark contrast between them. On the contrary, the simulated BF in Fig. 4(a2) shows interference features around Ca and O atomic columns, which cannot be resolved in the experimental BF image. However, both the ZLP [Fig. 4(a3)] and VP [Fig. 4(a4)] images can only identify the Nb and O atomic columns/interstitial sites with dark and bright contrast, respectively, while the visibility of Ca atomic columns remains challenging. Furthermore, the Nb atomic columns remain closely connected rather than well separated, similar to the experimental BF image in Fig. 4(a1). At the collection angle increased to 25 mrad, corresponding to $\alpha$, the separation of Nb atomic columns and the visibility of Ca atomic columns were not only improved, but also the bright contrast at O atomic columns or interstitial sites was reduced [see Fig. 4(b1)]. Similar to the recorded BF image, the simulated BF image in Fig. 4(b2) shows a very uniform bright contrast for both O atomic columns and interstitial sites, with no distinguishable features. Additionally, both ZLP [Fig. 4(b3)] and VP [Fig. 4(b4)] images can reveal Nb and Ca atomic columns with different contrasts. When the collection angle increases further to 50 mrad, the separation of neighboring Nb atomic columns becomes clearer, and the contrast arising from the Nb and Ca atomic columns considerably enhances [see Fig. 4(c1)] due to a reduction in contrast from the O atomic columns and interstitial sites. Because imaging at a collection angle of 50 mrad captures high-angle Kikuchi bands and the major contributions from Nb and Ca atoms, it also significantly reduces the contribution of O atoms. We will discuss this later in Figs. 6. This indicates that the interference of features decreased with increasing collection angles, resulting in a more uniform bright contrast. Interestingly, the BF images in Fig. 4(b1) and 4(c1) show contrast reversal compared to the HAADF image [see Fig. 3(b)]. In fact, the simulated BF image in Fig. 4(c2) closely matches the experimental BF image, except for the extremely bright contrast at O atomic columns and interstitial sites. Similarly, the contrast in the ZLP [Fig. 4(c3)] and VP [Fig. 4(c4)] images display the same contrast at atomic resolution. Therefore, atomic-resolution ZLP and VP images can be obtained when the EELS collection $\beta$ angle is equal to [19] or larger than the probe's convergence semi-angle ($\alpha$). These setups are used for atomic-resolution core-level elemental mapping [10], but our study shows that the ZLP images also exhibit contrast reversals similar to those in the experimental HABF images.

Using the method proposed by Zhu et al. [20] to remove artifacts associated with the elastic scattering of the electron probe, we corrected atomic-resolution VEELS images with the incoherent bright-field (IBF) image. Figures 5(a), 5(b), 5(c1), 5(c2), and 5(c3) show the recorded HAADF image, EEL spectrum, ZLP image, IBF image, and VP image, respectively. Here, the IBF image is the total low-loss image, which integrates energy loss from 1 to 60 eV. Elastically-corrected VP images at 15 ± 1 eV energy loss obtained from the STEM-EELS SI dataset are shown in Fig. 5(d1) for the collection angle $\beta = \alpha$, and in Fig. 5(d2) for $\beta > \alpha$, respectively. Thus, the elastic scattering-related atomic arrangements were essentially removed, leaving the delocalized signals unaffected. It should be noted, however, that the elastic-corrected VP images reported in ref. [20] still display inelastic signal confined within the $BaTiO_3$ and $SrTiO_3$ layers, with a delocalization length of 0.2 to 0.4 nm, which is shorter than the 1.4 nm calculated using the delocalized formula $L_{50} \approx 0.5\lambda/(\theta_E^{3/4})$. This indicated that the inelastic scattering signals remain confined within crystalline materials, even when elastic contrast is present.

To further reduce strong forward electron scattering and the overlap of Bragg reflection discs, which cause the intense ZLP in atomic-resolution VEELS, we also used a weak-beam geometry, also known as the off-axis EELS setting [11,12]. This was achieved by moving the Ronchigram disc outside the entrance aperture and using large-angle Kikuchi bands as a source. Figure 6(a) shows an experimental CBED pattern along the same [001] zone axis, with an $\alpha$ of 25 mrad. Because the Dectris ELA detector has a 1028 x 512 pixel array, we recorded the central CBED pattern separately with a shorter acquisition time (about 2 seconds), then shifted the pattern to capture the high-angle Kikuchi bands with a longer acquisition time (about 50 seconds). The final combined 2-fold symmetry CBED pattern is shown in Fig. 6(a). The use of high-angle Kikuchi bands to detect atomic-column-dependent differences with a 70 pm electron probe was proposed earlier [40,41]. Using this method, we simulated atomic-column-dependent CBED patterns with the xHREM™ software (HREM Research Inc., Japan). The results for the Nb, Ca, and O atomic columns are shown in Figs. 6(b1)-(b3), respectively. Distinct Kikuchi bands appear at the Nb and Ca atomic columns [see Fig. 6(b2) and (b2)]. In contrast, the O atomic column only shows a diffuse background without Kikuchi bands [see Fig. 6(b3)], with the band intensity decreasing in the order: Nb column > Ca column >> O column. This suggests that the Kikuchi bands originate from TDS localized on the Nb and Ca atomic columns. HAADF images can be generated by inserting an ADF detector [see the blue dashed circle representing the inner angle of the ADF detector in Figs. 6(b1)-(b3)]. Additionally, we visualized TDS signals by selecting the high-angle Kikuchi band as the electron source for weak-beam (off-axis) EELS imaging.

Based on the geometry between the probe's convergent semi-angle ($\alpha$) and the ADF collection angle, we carefully shifted the CBED pattern to select the horizontal Kikuchi band corresponding to {001} planes for entry into the EELS entrance aperture [see inset in Fig. 6(c1)]. We also aimed to prevent the central Ronchigram disc from touching the inner ADF detector, as this could alter ADF image contrast. Weak-beam (off-axis) HAADF, ZLP, and VP images are shown in Fig. 6(d1), 6(d2), and 6(d3), respectively. Both the weak-beam (off-axis) ZLP and VP images revealed atomic-resolution contrast similar to that of the HAADF image. This was expected because the selected {001} Kikuchi band is influenced by TDS localized on the Nb column (major) and the Ca columns [see Fig. 6(b1) and 6(b2)], illustrating a contrast reversal relative to the on-axis setup. The contrast between the ZLP image and the VP image was the same, unlike the contrast reversal in atomic-scale phonon spectroscopic imaging between the ZLP image and the phonon map [42]. Interestingly, VEEL spectra can still be obtained under this weak-beam (off-axis) geometry, even though the scattering angle exceeds 30 mrad, which is much larger than the inelastic scattering angles ($\theta_E$ is up to 0.125 mrad) in the VEELS region. In fact, for the same acquisition time, both weak-beam (off-axis) ZLP and VEELS intensities [see Fig. 6(c1)-(c4)] were not only significantly suppressed but also decreased by about 40 to 90 times compared to the on-axis setup. Unlike for the on-axis setup, there were noticeable decreases in intensity for both ZLP and VEELS intensities in the following sequence: the Nb column > the Ca column > the O column between two Nb layers > the interstitial site [see in Figs. 6(c1) to 6(c4)]. This trend was consistent with intensity variations observed in the ZLP and VP images shown in Figs. 6(d2) and 6(d3). In this way, by using the 70 pm electron probe positioned at the same atomic sites (Nb, Ca, O, and interstitial positions), we demonstrated atomic-resolution position-dependent ZLP and VEELS images, as shown in Figs. 3-6.

Figures 7(a1) and 7(a2) present probe position-dependent VEELS spectra over the energy-loss range from 2.5 to 4.5 eV to illustrate the additional broadened feature displayed in Fig. 2(b). A shoulder at about 3.3 to 3.4 eV energy loss indicates significant oscillation strength at Nb atomic columns [blue fitting curve in Fig. 7 (a1)], while the interstitial sites show two features: one between 3.1 and 3.2 eV energy loss [red fitting curve in Fig. 7 (a2)], and the other between 3.3 and 3.4 eV energy loss [blue fitting curve in Fig. 7 (a2)]. This suggests that the two broad features correspond to two different excited states. To elucidate the potential excitation mechanism during electron channeling, we simulated an electron probe propagating along the Nb column of the [$NbO_6$] octahedra and through the interstitial sites between two Ca columns, as indicated by the cyan and gray circles in Fig. 7(b1). The simulated probe wave functions are plotted in Figs. 7(c1) and 7(c2) in orthogonal views along [100] (red) and [010] (blue). Brightness indicates the intensity of the probe wave function at each depth. When the electron probe is located on the Nb atomic columns, it remains tightly confined, meaning the incident electron is channeled straight along the atomic column. In contrast, the probe incident at the interstitial site spreads into the surrounding area after propagating more than 5 nm, a phenomenon called electron dechanneling [43], which then broadens due to crosstalk with neighboring Nb and Ca atomic columns. Therefore, the feature at 3.2 to 3.5 eV energy loss stems mainly from the Nb atomic columns associated with the [$NbO_6$] octahedra, while the same feature at interstitial sites results from crosstalk caused by electron-probe dechanneling. Figures 7(b1) and 7(b2) show the simultaneously acquired HAADF image [Fig. 7(b1)] and the VEELS image for the range from 3 to 3.5 eV energy loss [Fig. 7(b2)], corresponding to the interband transition discussed in Figs. 2(b) and 2(d)]. In this case, the [$NbO_6$] octahedral units display slightly higher intensity compared to the nearly dark Ca atomic columns.

The broad feature at an energy loss of 3.1 to 3.2 eV in Fig. 7(a2) appears when the electron probe is positioned at an interstitial site and then passes through it as a channel. In fact, the value of $\varepsilon_1$ is large enough [see the enlarged dielectric function $\varepsilon(\omega)= \varepsilon_1(\omega)+ i\varepsilon_2(\omega)$ inset in Fig. 7(c)] to enable Čerenkov radiation (CR), since the criterion of $\varepsilon_1 \cdot (v/c)^2 > 1$ is fulfilled [1,2,15,44,45], and the electron probe propagates through an atom-sized cylinder [45]. To elucidate this possibility, we calculated the relativistic energy versus momentum (*E–k* maps) for 35 nm thick CNO slabs using Kröger's equation [46]. The calculated *E–k* maps in Fig. 7(c) show that the CR peak between 3.1 and 3.2 eV energy loss exhibits dispersion features. Therefore, VEELS spectra of CNO crystals with a thickness less than 35 nm are likely to excite the CR mode, which can be observed as an energy-loss peak between 3.1 and 3.2 eV at interstitial sites.

In summary, we experimentally demonstrated atomically resolved on-axis and off-axis VEELS imaging of $CaNb_2O_6$ single crystals grown by the optical floating-zone method. The atomic column contrast with on-axis ZLP imaging using a 70 pm electron probe matched the HABF signal at a collection semi-angle ($\beta$) equal to or larger than the convergent semi-angle ($\alpha$), while showing contrast reversal compared to conventional HAADF signals. Additionally, on-axis VEELS imaging of both plexcitons at about 7.3 eV and VPs at 15 eV energy loss also displayed atomically resolved contrast similar to the HABF signal contrast, indicating a decrease in contrast with increasing atomic number $Z$. To account for elastic contrast preservation, we analyzed the collection-angle effect and minimized its influence to obtain the non-atomic-resolution, delocalized VP images. Finally, we showed that the weak-beam (off-axis) ZLP and VEELS images also exhibit atomic resolution, but with contrast reversal relative to on-axis VEELS imaging. Furthermore, we found that the $NbO_6$ octahedral units directly contributed to the lateral maps of interband transitions from 3.2 to 3.5 eV energy loss, which relate to the O-2*p* and Nb-4*d* states.

## ACKNOWLEDGEMENTS

GJS acknowledges the support provided by MOST-Taiwan under project number 114-2622-E-027-020. The authors also thank Prof. Cheng-Hsiang Chen (NTU), Arron C. Johnston-Peck and Andrew A. Herzing (NIST) for helpful discussions and comments.

## DATA AVAILABILITY

The data that support the findings of this article are not publicly available upon publication because it is not technically feasible and/or the cost of preparing, depositing, and hosting the data would be prohibitive within the terms of this research project. The data are available from the authors upon reasonable request.

## FIGURE CAPTIONS

**FIG. 1.** (Color online) (a) Atomic crystal structure of a $CaNb_2O_6$ crystal projected along the *c*-axis (a1) and *a*-axis (a2). (b) XRD pattern of $CaNb_2O_6$ powders pulverized from the $CaNb_2O_6$ single crystals. It also shows the simulated XRD patterns for columbite from ICSD card No. 15208 [26] (blue curve) and for Rynersonite from ICSD card No. 148209 [27] (red curve) for comparison. The inset shows a color photograph of the $CaNb_2O_6$ single crystal grown by an optical floating-zone method. (c) Energy-dispersive X-ray spectrum obtained from the specimen. The insets, from left to right, are the HAADF image and elemental maps of Ca, Nb, and O, in yellow, cyan, and red, respectively. The experimental SAED pattern of the $CaNb_2O_6$ single crystal was acquired along the [001] zone axis (d1), and along the [100] zone axis (e1). Corresponding atomically resolved experimental HAADF images at the [001] zone axis (d2), and along the [100] zone axis (e2). The inset in (d2) shows the crystallographic projection along [001], along with a simulated HAADF image by using the xHREM$^{TM}$ software package for a sample thickness of 35 nm and Debye-Waller (DW) factors listed in Table 1. The yellow dashed box indicates the unit cell. The inset in (e2) shows the simulated HAADF image.

**FIG. 2.** (Color online) *Aloof* beam valence EELS spectra collected from the $CaNb_2O_6$ single crystal at several positions of the electron probe positioned at the locations marked in the HAADF image in (c). The zero-loss peak was subtracted from all the spectra in (a), and the intensity was normalized to the spectral feature at 7 ± 1 eV, then vertically shifted for visualization. The orange curve shows the fitting used to assign the peak's energies. (b) Complex dielectric function [$\varepsilon(\omega)= \varepsilon_1(\omega)+ i\varepsilon_2(\omega)$] of $CaNb_2O_6$ derived from the purple spectrum in (a). The EELS spectrum in the inset illustrates the bandgap measurement. (c) From left to right is the HAADF image, and STEM-SI intensity maps acquired at the energy losses of 0 ± 1 eV, 7 ± 1 eV, and 15 ± 1 eV. (d) The calculated partial density of states (PDOS) for $CaNb_2O_6$.

**FIG. 3.** (Color online) Position-dependent zero-loss peak and VEELS spectra recorded from different atomic columns, marked in the HAADF image in (b). (a1) Nb atomic column. (a2) Ca atomic column. (a3) Oxygen (O) atomic column located between two Nb atomic layers. (a4) Interstitial site between two Ca atomic columns. The inset in (a1) shows the on-axis EELS geometry setup. (b) Simultaneously acquired HAADF image. The top right inset shows the crystallographic projection along [001]. (c1-c3) ZLP image (c1), low-loss images at 7 ± 1 eV (c2), and 15 ± 1 eV energy loss

(c3), respectively. Insets in (b) and (c1-c3) depict the schematic atomic arrangements of Ca, Nb, O, and interstitial sites, represented by orange, cyan, red, and gray, respectively.

**FIG. 4.** (Color online) (a1-a4) Experimental bright-field (BF) image (a1), simulated BF image (a2), ZLP image (a3), and VP image at 15 ± 1 eV energy loss (a4) for $\alpha$= 25 mrad, with a BF collection angle ranging from 0 to 20 mrad, and $\beta$= 20 mrad. (b1-b4) Experimental BF image (b1), simulated BF image (b2), ZLP image (b3), and VP image at 15 ± 1 eV energy loss (b4) for $\alpha$= 25 mrad, with a BF collection angle ranging from 0 to 25 mrad, and $\beta$= 25mrad. (c1-c4) Experimental BF image (c1), simulated BF image (c2), ZLP image (c3) and VP image at 15 ± 1 eV energy loss (c4) for $\alpha$= 25 mrad, with a BF collection angle ranging from 0 to 50 mrad, and $\beta$= 50 mrad. The scale bars in (a1), (b1), and (c1) are 0.5 nm. The inset shows the schematic atomic arrangements of Ca, Nb, and O, in orange, cyan, and red, respectively.

**FIG. 5.** (Color online) (a) Simultaneously recorded HAADF image. (b) Representative EEL low-loss spectrum. (c1) ZLP image integrated from the blue area at 0 ± 1 eV in (b). (c2) Total EELS signal image integrated from the gray area from 1 to 60 eV energy loss in (b). (c3) VP image integrated from the red area at 15 ± 1 eV energy loss in (b). (d1-d2) Elastically-corrected VP images at 15 ± 1 eV energy loss obtained from the STEM-EELS SI dataset recorded for $\beta$= $\alpha$ (d1) and for $\beta$> $\alpha$ (d2). EELS acquisition conditions were as follows: $\alpha$= 25 mrad, and $\beta$= 25 mrad in (b, c, d1), and $\beta$= 50 mrad in (d2).

**FIG. 6.** (Color online) (a) Experimental CBED pattern. (b1-b3) Simulated atomic-column-dependent CBED patterns: Nb (b1), Ca (b2), and O (b3), calculated using the xHREM$^{TM}$ software package for a sample thickness of 35 nm. The inset in (b1-b3) are the simulated HAADF image with marked electron-probe positions for the Nb (b1), Ca (b2), and O (b3) atomic columns. The blue dashed circle in (b1-b3) indicates the inner ADF collection angle. Orange circles in (b1-b3) highlight the differences and are used for off-axis EELS setup. (c1-c4) display position-dependent zero-loss peak and VEELS spectra recorded from the atomic columns indicated in the HAADF image in (d1) with an off-axis EELS setup. (c1) represents the Nb atomic column. (c2) represents the Ca atomic column. (c3) shows O atomic column between two Nb atomic layers. (c4) depicts the interstitial site between two Ca atomic columns. The inset in (c1) illustrates the off-axis EELS geometry setup. (d1-d3) show the simultaneously acquired HAADF image (d1), ZLP image (d2), and VP image at 15 ± 1 eV energy loss (d3).

**FIG. 7.** (Color online) Position-dependent VEELS spectra recorded from the Nb atomic column (a1) and interstitial site (a2) after removing the ZLP peak; focus on the energy-loss range of 2.5-4.5 eV.

Insets in (a1) and (a2) show the simulated wave function with orthogonal view along [100] (red) and [010] (blue). (b1-b2) Displays the simultaneously acquired HAADF image (b1) and low-loss image at 3.25 ± 0.25 eV (b2). (c) Shows the calculated relativistic *E*-*k* maps (E is energy loss and *k* is momentum transfer) for $CaNb_2O_6$ slabs with thicknesses of 35 nm at an accelerating voltage of 200 kV. The inset depicts the complex dielectric function [$\varepsilon(\omega)= \varepsilon_1(\omega)+ i\varepsilon_2(\omega)$] from 2.5 eV to 4.5 eV.

**TABLE 1** The refinement results of $CaNb_2O_6$.

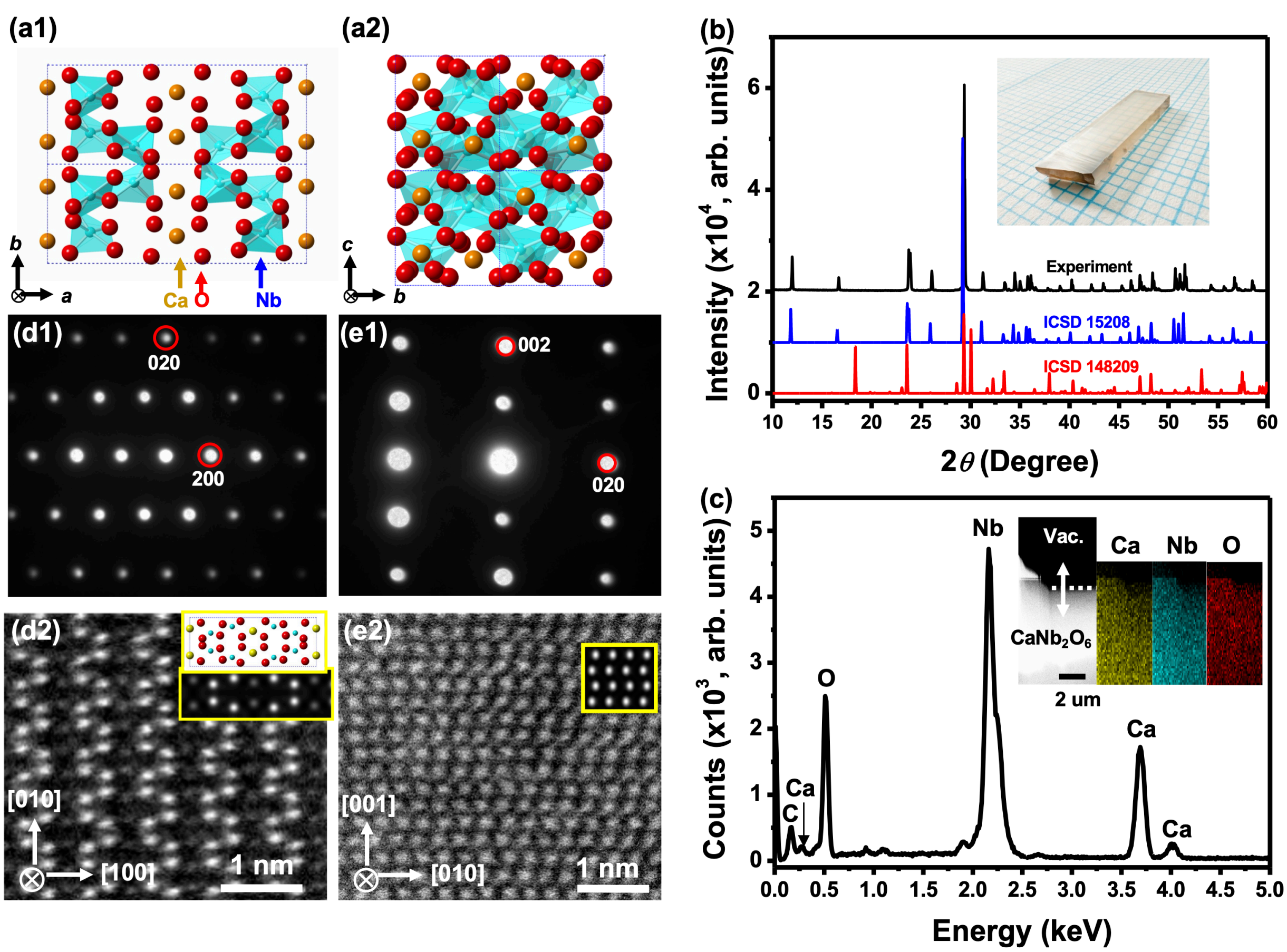
(a1)
(a2)
(b)
(c)
(d1)
(e1)
(d2)
(e2)
Ca
O
Nb
020
200
002
020
[010]
[100]
[001]
[010]
1 nm
1 nm
Experiment
ICSD 15208
ICSD 148209
Intensity (x10⁴, arb. units)
2θ (Degree)
Counts (x10³, arb. units)
Energy (keV)
Vac.
CaNb2O6
2 um
Nb
O
Ca
Ca
C
Ca

Figure 1

Table 1

| **Space group = *Pbcn* (No. 60)**<br>**a = 1.4981(2) nm, b = 0.5749(7) nm, c = 0.5224(6) nm**<br>**$R_{wp}$ = 8.36%** | | | | | |
|---|---|---|---|---|---|
| **Atom** | **Site** | **x** | **y** | **z** | **$B_{eq}$ / $nm^2$** |
| **Ca** | **4c** | **0** | **0.223(7)** | **0.75** | **34.99** |
| **Nb** | **8d** | **0.165(1)** | **0.316(2)** | **0.303(2)** | **16.91** |
| **O1** | **8d** | **0.086(6)** | **0.103(1)** | **0.401(2)** | **149.9** |
| **O2** | **8d** | **0.098(6)** | **0.426(2)** | **0.008(2)** | **184.3** |
| **O3** | **8d** | **0.260(6)** | **0.128(2)** | **0.129(2)** | **46.38** |

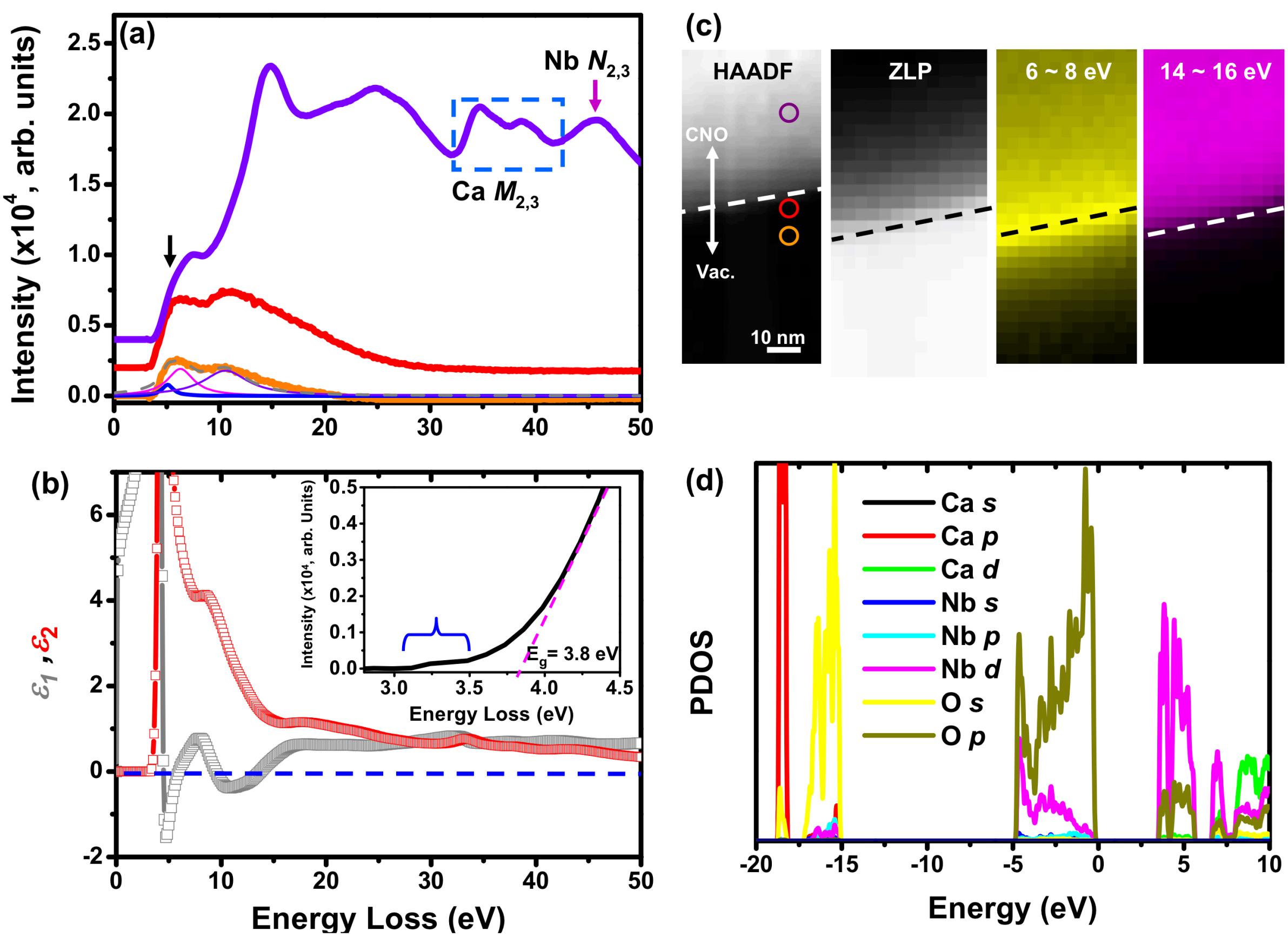
(a)
Intensity (x10⁴, arb. units)
Nb $N_{2,3}$
Ca $M_{2,3}$
(b)
$\varepsilon_1$, $\varepsilon_2$
Intensity (x10⁴, arb. Units)
$E_g$ = 3.8 eV
Energy Loss (eV)
Energy Loss (eV)
(c)
HAADF
ZLP
6 ~ 8 eV
14 ~ 16 eV
CNO
Vac.
10 nm
(d)
Ca s
Ca p
Ca d
Nb s
Nb p
Nb d
O s
O p
PDOS
Energy (eV)

Figure 2

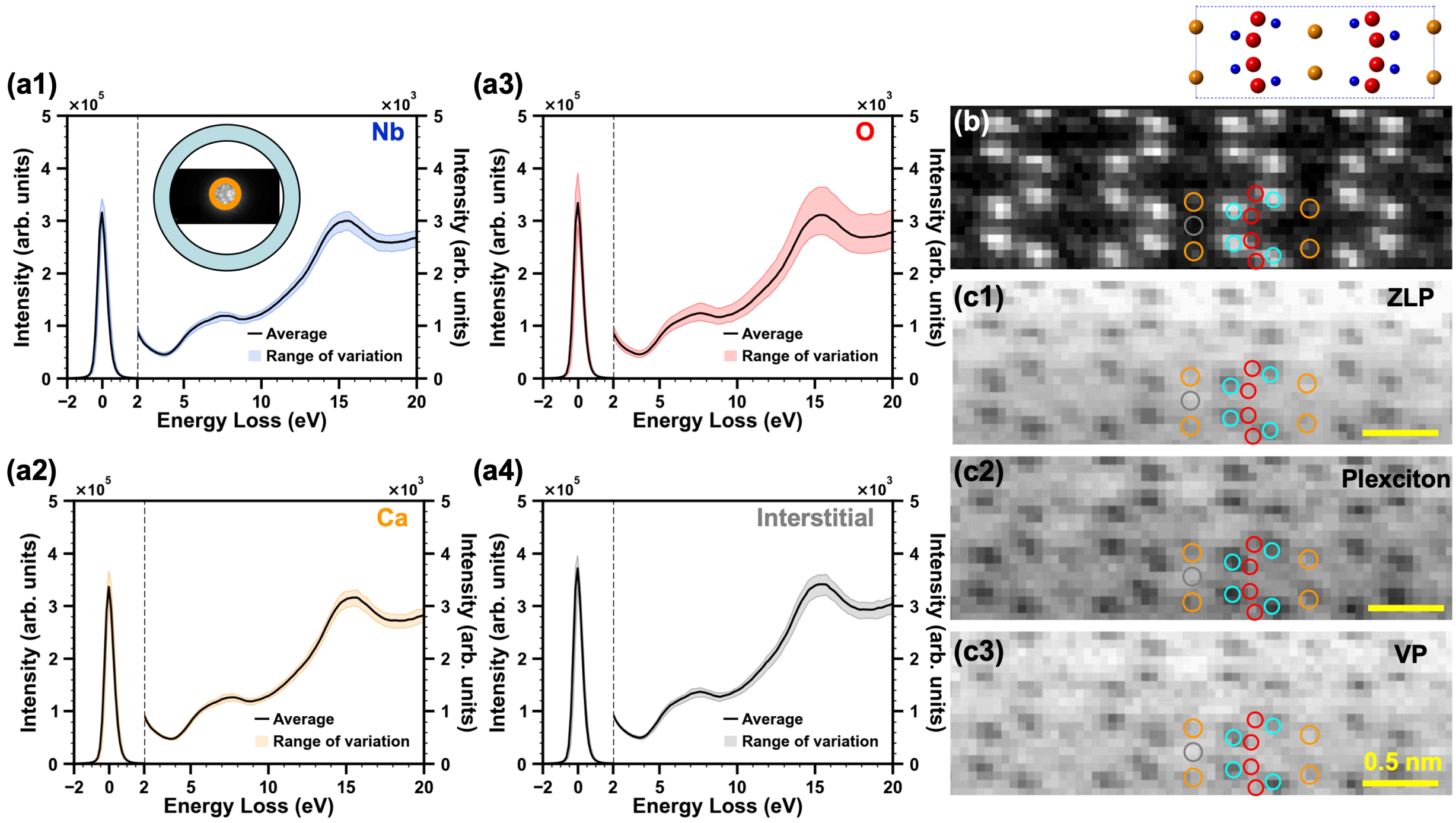
(a1)
Nb
Average
Range of variation
Energy Loss (eV)
Intensity (arb. units)
(a2)
Ca
(a3)
O
(a4)
Interstitial
(b)
(c1)
ZLP
(c2)
Plexciton
(c3)
VP
0.5 nm

Figure 3

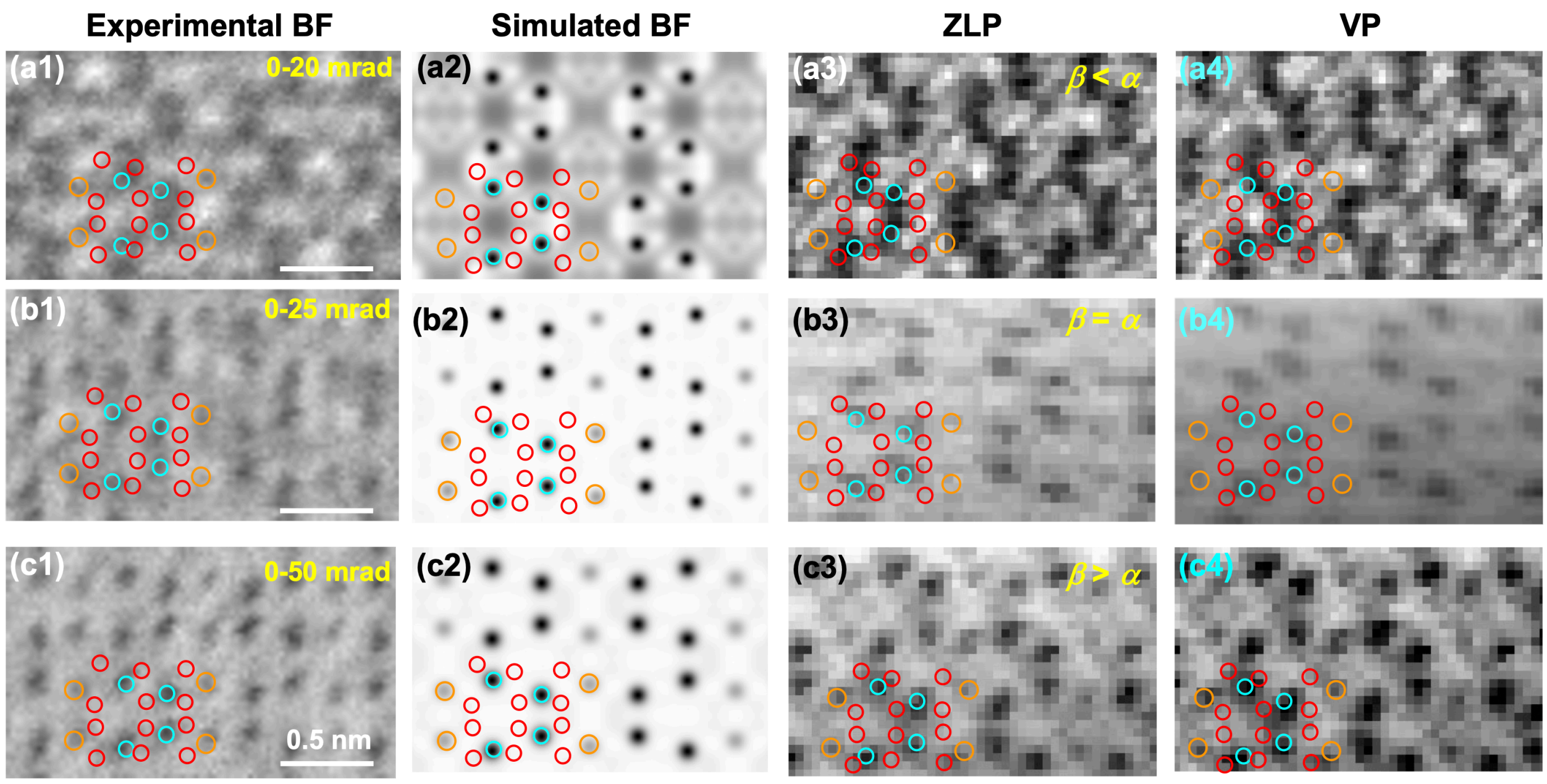
Experimental BF
Simulated BF
ZLP
VP
(a1)
0-20 mrad
(a2)
(a3)
β < α
(a4)
(b1)
0-25 mrad
(b2)
(b3)
β = α
(b4)
(c1)
0-50 mrad
0.5 nm
(c2)
(c3)
β > α
(c4)

Figure 4

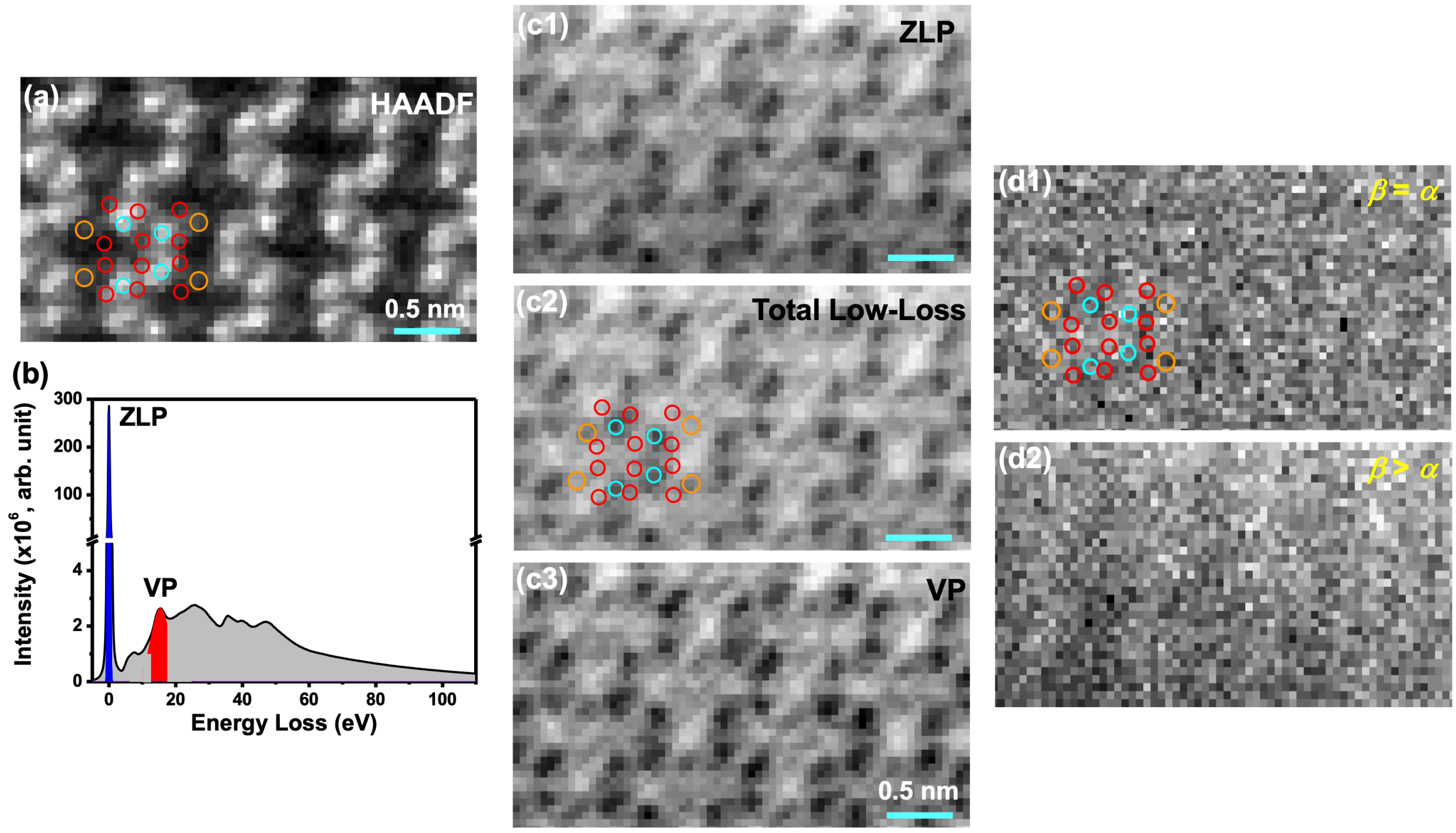
(a)
HAADF
0.5 nm
(b)
ZLP
VP
Intensity (x10^6, arb. unit)
Energy Loss (eV)
(c1)
ZLP
(c2)
Total Low-Loss
(c3)
VP
0.5 nm
(d1)
β = α
(d2)
β > α

Figure 5

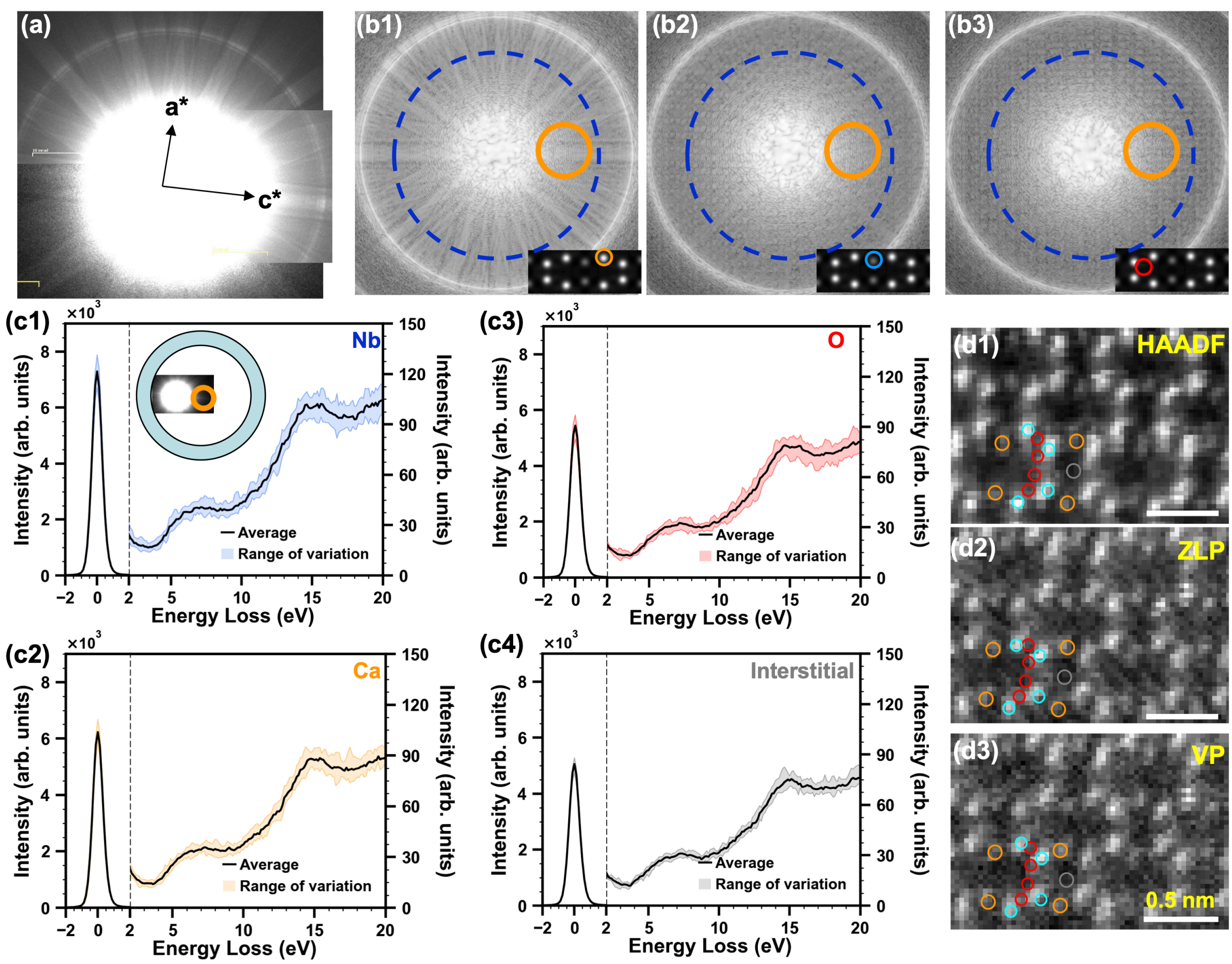

(a)
a*
c*
(b1)
(b2)
(b3)
(c1)
Nb
Average
Range of variation
Intensity (arb. units)
Energy Loss (eV)
(c2)
Ca
(c3)
O
(c4)
Interstitial
(d1)
HAADF
(d2)
ZLP
(d3)
VP
0.5 nm

Figure 6

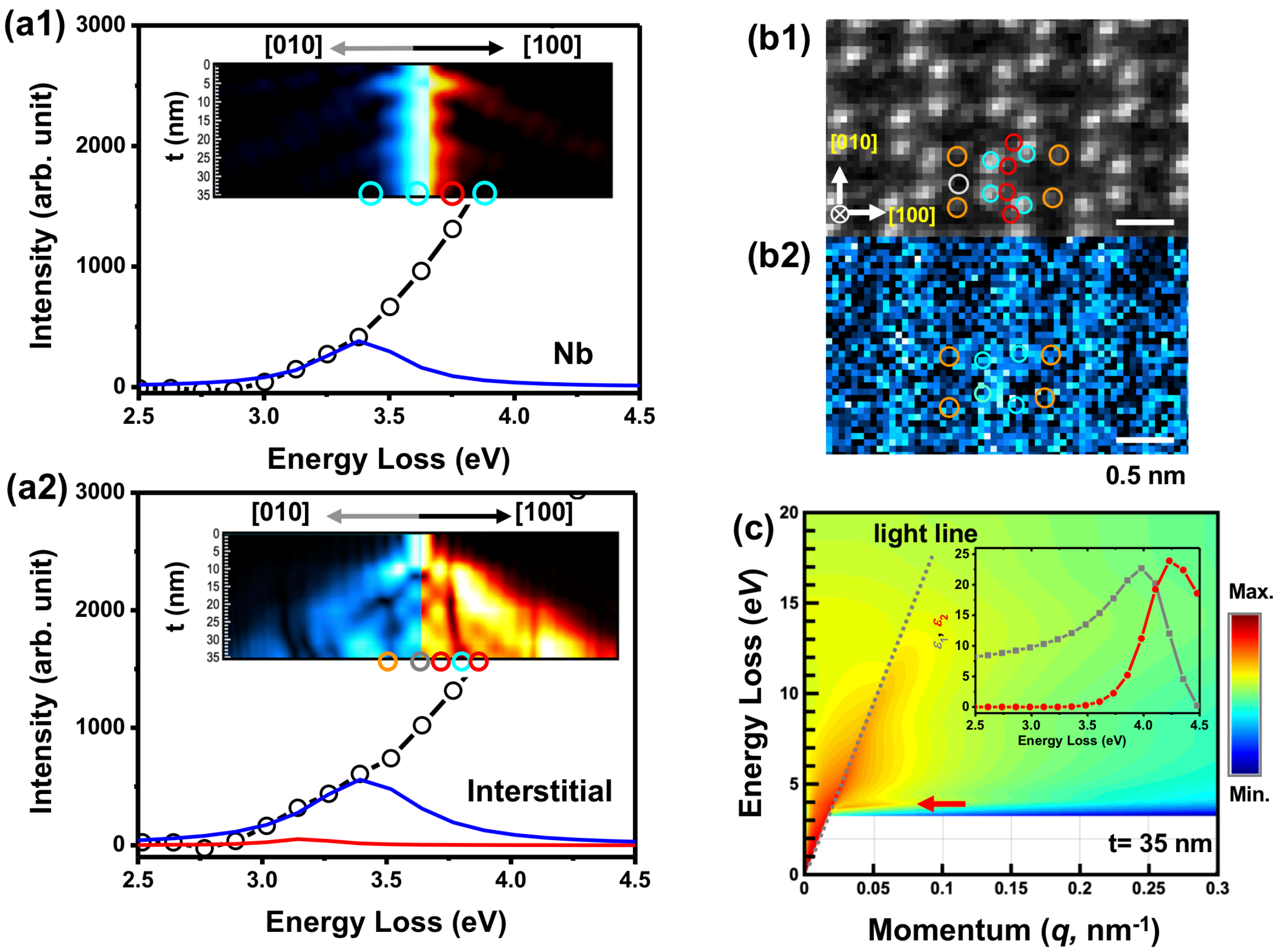
(a1)
[010]
[100]
t (nm)
Intensity (arb. unit)
Nb
Energy Loss (eV)
(a2)
Interstitial
(b1)
(b2)
0.5 nm
(c)
light line
Energy Loss (eV)
Energy Loss (eV)
Max.
Min.
t= 35 nm
Momentum (q, nm-1)

Figure 7